\documentclass[ twocolumn,aps,prd,   
               preprintnumbers,numbers,sort&compress,nofootinbib,
                            showpacs,
               colorlinks,
               linkcolor=blue,
               citecolor=blue]{revtex4-1}
\newcommand{\exclude}[1]{}

\usepackage{graphicx,amsmath,amssymb,bm}
\usepackage{psfrag}
\usepackage{feynmp}
\usepackage{hyperref}
\usepackage{enumitem}

\newcommand{\beq}{\begin{equation}}
\newcommand{\eeq}{\end{equation}}
\newcommand{\be}{\begin{eqnarray}}
\newcommand{\ee}{\end{eqnarray}}

\def\+{\dagger}
\def\la{\langle}
\def\ra{\rangle}
\def\<{\langle}
\def\>{\rangle}
\def\mb{\mathbf}

\usepackage{color}

\begin{document}

\title{Long range order in gauge theories. Deformed QCD as a toy model.}

\author{Evan Thomas \& Ariel R. Zhitnitsky} 
\affiliation{Department of Physics \& Astronomy, University of British Columbia, Vancouver, B.C. V6T 1Z1, Canada}

\begin{abstract}
\exclude{
We study a long range order in a ``deformed QCD'' model, which although weakly coupled, exhibits the important topological structure of QCD.
Motivated by the observation of coherent low-dimensional vacuum configurations in Monte-Carlo lattice simulations, we consider domain wall solutions and their interaction with localized topological ``monopoles''.
We then show a net attractive interaction between the two, such that a picture of all localized topological charge becoming bound to large extended objects (surfaces) emerges. }

We study a number of different ingredients, related to long range order observed  in lattice QCD simulations, using a simple ``deformed QCD'' model. 
This model is a weakly coupled gauge theory, which however has all the relevant crucial elements allowing us to study difficult and nontrivial problems which are known to be present in real strongly coupled QCD. 
In the present study, we want to understand the physics of long range order in form of coherent low dimensional vacuum configurations observed in Monte Carlo lattice simulations.
We demonstrate the presence of double-layer domain wall structures in the deformed QCD, and study their interaction with localized topological monopoles.
Furthermore, we show that there is in fact an attractive interaction between the two, such that the monopole favors a position within the domain wall.

\end{abstract}

\maketitle

\section{Introduction and motivation} \label{introduction}

The main motivation for this work is the recent Monte Carlo studies of pure glue gauge theory which have revealed some very unusual features. To be more specific, the gauge configurations display a laminar structure in the vacuum consisting of extended, thin, coherent, locally low-dimensional sheets of topological charge embedded in 4d space, with opposite sign sheets interleaved, see original QCD lattice results~\cite{Horvath:2003yj,Horvath:2005rv,Horvath:2005cv,Alexandru:2005bn}.
A similar structure has been also observed in QCD by different groups~\cite{Ilgenfritz:2007xu,Ilgenfritz:2008ia,Bruckmann:2011ve,Kovalenko:2004xm,Buividovich:2011cv} and also in two dimensional $CP^{N-1}$ model~\cite{Ahmad:2005dr}.
Furthermore, the studies of localization properties of Dirac eigenmodes have also shown evidence for the delocalization of low-lying modes on effectively low-dimensional surfaces.
The following is a list of the key properties of these gauge configurations which we wish to study:

1) The tension of the ``low dimensional objects''  vanishes below the critical temperature and these objects percolate through the vacuum, forming a kind of a vacuum condensate; 

2) These ``objects'' do not percolate through the whole 4d volume, but rather, lie on low dimensional surfaces $1\leq d < 4$ which   organize  a coherent   double layer structure; 

3) The total area of the surfaces is dominated by a single percolating cluster of ``low dimensional object''; 

4) The contribution of the percolating objects to the topological susceptibility has the same sign compared to its total value; 

5) The width of the percolating objects apparently vanishes in the continuum limit; 

6) The density of well localized 4d objects (such as small size instantons) apparently vanishes in the continuum limit.

It is very difficult to understand the above properties using conventional quantum field theory analysis. 
Indeed, the QCD lattice results ~\cite{Horvath:2003yj,Horvath:2005rv,Horvath:2005cv,Alexandru:2005bn,Ilgenfritz:2007xu,Ilgenfritz:2008ia,Bruckmann:2011ve,Kovalenko:2004xm} imply that the topological density distribution is not localized in any finite size configurations such as instantons; rather the topological density is spread out on the surface of low-dimensional sheets.
Such a structure can not be immediately seen in gluodynamics, at least not at the semiclassical level.  
At the same time, these Monte Carlo results could be interpreted very nicely with a conjecture that the observed structure is identified with the extended D2 branes in holographic description\cite{Gorsky:2007bi,Gorsky:2009me,Zhitnitsky:2011aa,BKYZ}.

One of the key elements of this conjecture is assumption that  the tension of the D2 branes vanishes below the QCD phase transition $T<T_c$ such that an arbitrary large number of these objects can be formed.
Vanishing tension in the dual description in the  confined phase is a result of the Hawking-Page phase transition~\cite{wittenterm}  when the  D2 brane shrinks to the tip of a cigar type geometry.

 The second key element in identification of the structure observed on the lattice ~\cite{Horvath:2003yj,Horvath:2005rv,Horvath:2005cv,Alexandru:2005bn,Ilgenfritz:2007xu,Ilgenfritz:2008ia,Bruckmann:2011ve,Kovalenko:2004xm} with the holographic description in terms of the D branes is the assumption that the topological density distribution which is originally localized in well defined D0 branes (instantons),  somehow spreads out along extended D2 branes as a result of strong interaction between  D0-D2  branes, leading  to their binding.
Such a picture  was basically motivated, as mentioned in \cite{Gorsky:2009me,Zhitnitsky:2011aa,BKYZ}, by the structure which emerges in supersymmetric field theories~\cite{Davies:1999uw} where the relevant dynamics can be indeed formulated in terms of the strongly bound D0-D2 configurations.

In this paper, we investigate precisely the second idea above in the framework of a ``deformed QCD'' developed in \cite{Yaffe:2008}. 
The deformation allows us to bring the gauge theory into a weakly coupled regime wherein calculations can be performed in theoretically controllable manner.
In spite of the great deal of analytic control provided, the deformed theory preserves many of the relevant structures present in strongly coupled QCD including confinement, degeneracy of topological sectors, and the correct nontrivial $\theta$ dependence.
Furthermore, it seems, there is no order parameter differentiating the weakly coupled deformed regime from the strongly coupled regime, which reproduces undeformed QCD \cite{Yaffe:2008}, so that the (gross) behaviour of the two theories may be quite similar.

In particular, the deformed theory exhibits  two important structures of note: first, the topological charge in this model is carried by the fractionally charged monopoles with topological charges $Q=\pm 1/N$; and second, there are domain walls present in the system as a result of a generic $2\pi$ periodicity of the effective low energy Lagrangian governing the dynamics. Given these ingredients, we would like to test the following two ideas which are apparently related to the  configurations observed  in the  lattice simulations ~\cite{Horvath:2003yj,Horvath:2005rv,Horvath:2005cv,Alexandru:2005bn,Ilgenfritz:2007xu,Ilgenfritz:2008ia,Bruckmann:2011ve,Kovalenko:2004xm}: 

1) the domain walls form precisely a double layer structure with opposite sign sheets of the topological charge density interleaved;  

2) the monopoles and domain walls attract each other and the topological charge originally localized on monopoles spreads out along the domain walls. 

If the second occurs, there will be few well-localized finite sized sources carrying the topological charge. Instead, the topological charge density will be spread over extended domain walls, which is precisely what has been observed in simulations ~\cite{Horvath:2003yj,Horvath:2005rv,Horvath:2005cv,Alexandru:2005bn,Ilgenfritz:2007xu,Ilgenfritz:2008ia,Bruckmann:2011ve,Kovalenko:2004xm}. 

We note that a similar picture of attraction between monopoles and domain walls was originally discussed in a cosmological context \cite{Dvali:1997sa}, see also the related papers \cite{Dvali:1996xe,Alexander:2000yx,Dvali:2002fi} and references therein.
The basic idea in \cite{Dvali:1997sa} is that if physical monopoles and domain walls are present in the system, there will be an attractive force between them.
Then, if these objects collide, the monopole's winding number (monopole charge) spreads out on the surface of the domain wall, and will be eventually pushed to the boundaries at infinity.
This effect was suggested as a solution of the so-called ``cosmological monopole's problem''. 
In our context we do not have real physical monopoles and real physical domain walls in Minkowski space, but rather Euclidean monopoles and domain walls which must be interpreted as configurations describing the tunnelling processes in physical Minkowski space, see detail discussions of this point in \cite{Zhitnitsky:2011aa}. 
\exclude{as corresponding multiple vacua are not distinct physical vacuum states but rather the states which can be related by large gauge transformations} 
Nevertheless, the formal structure of the problem and relevant features (such as attraction between the objects and spreading the magnetic charge over the surface) are very much the same.

We also note that ``deformed QCD'' model has been successfully used to test some other nontrivial features of strongly coupled QCD such as emergence of non-dispersive contact term in the topological susceptibility \cite{Thomas:2011ee} and the emergence of a topological Casimir behaviour in gauge theory with a gap \cite{Thomas:2012ib}.
In both cases the effects of interest are a result of the nontrivial topological vacuum structure of this model.

The structure of our presentation is as follows. 
In next section (\ref{deformedqcd}), we review the relevant parts of the model \cite{Yaffe:2008,Thomas:2011ee} including the low energy description of the theory in terms of the sine-Gordon Lagrangian.
In the following section (\ref{dw}), we construct the domain walls and explicitly demonstrate the double layer structure apparently observed on the lattices.  Finally, in Section \ref{dw-monopoles} we study the interaction of the domain walls and monopoles.

\exclude{and shows how magnetic and topological charges can be spreading along the domain walls, which is also  apparently observed on the lattices.}

\section{Deformed QCD} \label{deformedqcd}

In the ``deformed'' Yang-Mills, developed in \cite{Yaffe:2008}, an extra ``center-stabilization'' term is put into the the Lagrangian in order to prevent the center symmetry breaking that characterizes the QCD phase transition between ``confined'' hadronic matter and ``deconfined'' quark-gluon plasma.
Thus we have a theory which remains confined at high temperature in a weak coupling regime, and for which it is claimed \cite{Yaffe:2008} that there does not exist an order parameter to differentiate the low temperature (non-abelian) confined regime from the high temperature (abelian) confined regime.
We now proceed, in section \ref{model}, to review the relevant aspects of the theory.
We then discuss, in section \ref{lagrangian}, the low-energy effective Lagrangian which gives rise to the domain wall solutions mentioned earlier.

\subsection{Formulation of the theory}\label{model}

We start with pure $SU(N)$ Yang-Mills (gluodynamics) Wick rotated with Euclidean time compactified on the manifold $\mathbb{R}^{3} \times S^{1}$ defined by the standard action
\be \label{standardYM}
  S^{YM} = \int_{\mathbb{R}^{3} \times S^{1}} d^{4}x\; \frac{1}{2 g^2} \mathrm{tr} \left[ F_{\mu\nu}^{2} (x) \right].
\ee
We then add to it a deformation action,
\be \label{deformation}
  \Delta S \equiv \int_{\mathbb{R}^{3}}d^{3}x \; \frac{1}{L^{3}} P \left[ \Omega(\mathbf{x}) \right],
\ee 
built out of the Wilson loop (Polyakov loop) wrapping the compact dimension,
\be \label{loop}
  \Omega(\mathbf{x}) \equiv \mathcal{P} \left[ e^{i \oint dx_{4} \; A_{4} (\mathbf{x},x_{4})} \right],
\ee
where $L$ is the length of the compact time dimension.
The ``double-trace" deformation potential $P \left[ \Omega \right]$ respects the symmetries of the original theory and is built to stabilize the phase with unbroken center symmetry.
It is defined by
\be \label{deformationpotential}
  P \left[ \Omega \right] \equiv \sum_{n = 1}^{\lfloor N/2 \rfloor} a_{n}
    \left| \mathrm{tr} \left[ \Omega^{n} \right] \right| ^{2}.
\ee
Here $\lfloor N/2 \rfloor$ denotes the integer part of $N/2$ and $\left\{ a_{n} \right\}$ is a set of suitably large positive coefficients.

In the undeformed theory the effective potential for the Wilson loop is minimized for $\Omega$ an element of $\mathbb{{Z}}_{N}$.
The deformation potential (\ref{deformationpotential}) with sufficiently large $\{ a_{n} \}$ however changes the effective potential for the Wilson line so that it is minimized instead by configurations in which $\mathrm{tr} \left[ \Omega^{n} \right] = 0$, which in turn implies that the eigenvalues of $\Omega$ are uniformly distributed around the unit circle.
Thus, the set of eigenvalues is invariant under the $\mathbb{{Z}}_{N}$ transformations, which multiply each eigenvalue by $e^{2 \pi i k / N}$ (rotate the unit circle by $k/N$).
The center symmetry is then unbroken by construction.
The coefficients, $\{ a_{n} \}$, can be suitably chosen such that the deformation potential, $P\left[ \Omega \right]$, forces unbroken symmetry at any compactification scale \cite{Yaffe:2008}, but for our purposes we are only interested in small compactifications ($L \ll \Lambda^{-1}$ where $L$ is again the length of the compactified dimension and $\Lambda$ is the QCD scale).
At small compactification, the gauge coupling at the compactification scale is small so that the semiclassical computations are under complete theoretical control \cite{Yaffe:2008}.

\subsection{Infrared description}\label{lagrangian}
 
As discussed in \cite{Yaffe:2008}, the proper infrared description (at distances larger than the compactification scale) of the theory is a dilute gas of $N$ types of monopoles, characterized by their magnetic charges, which are proportional to the simple roots and affine root of the Lie algebra for the gauge group $U(1)^{N}$.
The extended root system is given by the simple roots,
\begin{equation}	\label{roots}
  \begin{array}{lclcl}
    \alpha_{1} &=& \left( 1, -1, 0, \dots, 0 \right) &=& \hat{e}_{1} - \hat{e}_{2},\\
    \alpha_{2} &=& \left( 0, 1, -1, \dots, 0 \right) &=& \hat{e}_{2} - \hat{e}_{3},\\
      &\vdots&	&  &  \\
    \alpha_{N-1} &=& \left( 0, \dots, 0, 1, -1 \right) &=& \hat{e}_{N-1} - \hat{e}_{N},
  \end{array}
\end{equation}
and the affine root,
\begin{equation*}
  \begin{array}{lclcl}
    \alpha_{N} &=& \left( -1, 0, \dots, 0 ,1 \right) &=& \hat{e}_{N} - \hat{e}_{1}. \\
  \end{array}
\end{equation*}
We denote this root system by $\Delta_{\mathrm{aff}}$ and note that the roots obey the inner product relation
\be \label{dotproduct}
  \alpha_{a} \cdot \alpha_{b} = 2\delta_{a, b} - \delta_{a, b+1} - \delta_{a, b-1}.
\ee

For a fundamental monopole with magnetic charge $\alpha_{a} \in \Delta_{\mathrm{aff}}$, the topological charge is given by
\be \label{topologicalcharge}
  Q = \int_{\mathbb{R}^{3} \times S^{1}} d^{4}x \; \frac{1}{16 \pi^{2}}
    \mathrm{tr} \left[ F_{\mu\nu} \tilde{F}^{\mu\nu} \right] = \pm\frac{1}{N},
\ee
and the Yang-Mills action is given by
\be \label{YMaction}
  S_{YM} &=& \int_{\mathbb{R}^{3} \times S^{1}} d^{4}x \;
    \frac{1}{2 g^{2}} \mathrm{tr} \left[ F_{\mu\nu}^{2} \right]\\
  &=& \left| \int_{\mathbb{R}^{3} \times S^{1}} d^{4}x \;
    \frac{1}{2 g^{2}} \mathrm{tr} \left[ F_{\mu\nu} \tilde{F}^{\mu\nu} \right] \right|
    = \frac{8 \pi^{2}}{g^{2}} \left| Q \right|. \nonumber
\ee
The second equivalence hold because the classical monopole solutions are self dual, $F_{\mu\nu} = \pm\tilde{F}_{\mu\nu}$.
 
The  $\theta$-parameter in the Yang-Mills action can be included in conventional way,
\be \label{thetaincluded}
  S_{\mathrm{YM}} \rightarrow S_{\mathrm{YM}} + i \theta \int_{\mathbb{R}^{3} \times S^{1}}
    d^{4}x\frac{1}{16 \pi^{2}} \mathrm{tr} \left[ F_{\mu\nu} \tilde{F}^{\mu\nu} \right],
\ee
with $\tilde{F}^{\mu\nu} \equiv \epsilon^{\mu\nu\rho\sigma} F_{\rho\sigma}$.

The system of interacting monopoles, including $\theta$ parameter, can be represented in the dual sine-Gordon form as follows \cite{Yaffe:2008,Thomas:2011ee},
\be
\label{thetaaction}
  S_{\mathrm{dual}} &=& \int_{\mathbb{R}^{3}}  d^{3}x \frac{1}{2 L} \left( \frac{g}{2 \pi} \right)^{2}
      \left( \nabla \bm{\sigma} \right)^{2} \nonumber \\
    &-& \zeta  \int_{\mathbb{R}^{3}}  d^{3}x \sum_{a = 1}^{N} 
      \cos \left( \alpha_{a} \cdot \bm{\sigma} + \frac{\theta}{N} \right),
\ee
where $\zeta$ is magnetic monopole fugacity which can be explicitly computed in this model using the conventional semiclassical approximation.
The $\theta$ parameter enters the effective Lagrangian (\ref{thetaaction}) as $\theta/N$ which is the direct consequence of the fractional topological charges of the monopoles (\ref{topologicalcharge}).
Nevertheless, the theory is still $2\pi$ periodic, but not because of an explicit $2\pi$ periodicity of Lagrangian (\ref{thetaaction}).
Rather, it is restored as a result of summation over all branches of the theory when the levels cross at $\theta=\pi ~(mod ~2\pi)$ and one branch replaces another and becomes the lowest energy state as discussed in \cite{Thomas:2011ee}.

\exclude{
Indeed, the ground state energy density is determined by minimization of the effective potential (\ref{thetaaction}) when summation over all branches is assumed in the definition of the canonical partition function. It  is given by
\be \label{min}
  E_{min} (\theta)=  - \lim_{V \rightarrow \infty} \; \frac{1}{VL} \ln 
    \left[\sum_{l=0}^{N-1}e^{ V\zeta N\cos \left( \frac{ \theta + 2 \pi \, l }{N} \right)}\right]
\ee
where $V$ is 3d volume of the system.
Equation (\ref{min}) shows that in the limit $V\rightarrow\infty$ cusp singularities occur at the values at $\theta=\pi ~(mod ~2\pi)$    where the lowest energy vacuum state switches from one analytic branch to another one.     
Such a pattern is known to emerge in many four dimensional supersymmetric models, and also gluodynamics in the limit $N=\infty$ \cite{witten_N}.
It has been further argued \cite{Halperin:1997bs} that the same  pattern also emerges in four dimensional gluodynamics at any finite $N$.
The same pattern emerges in holographic description of QCD ~\cite{wittenflux} at $N=\infty$ as well.
}
   
In the following sections we shall need an explicit expression for the topological density and magnetic field in terms of scalar $\bm{\sigma}$ field,
\be \label{topdensity}
  q(\mathbf{x}) & = & \displaystyle  \frac{1}{16 \pi^2} \mathrm{tr} \left[ F_{\mu\nu} \tilde{F}^{\mu\nu} \right] 
      =  \displaystyle \frac{-1}{8 \pi^2} \epsilon^{i j k 4} \sum_{a = 1}^{N} F^{(a)}_{j k} F^{(a)}_{i 4} \nonumber \\
    & = & \displaystyle \frac{g }{4 \pi^2} \sum_{a = 1}^{N} \left< A_{4}^{(a)} \right>
      \left[ \nabla \cdot \mathbf{B}^{(a)} (\mathbf{x}) \right],
\ee
where the $U(1)^N$ magnetic field, $B^i = \epsilon^{ijk4} F_{jk}/2g$ is expressed in terms of scalar magnetic potential as follows
\be \label{magnetic}
  F_{ij}^{(a)}=\frac{g^2}{2\pi L}\epsilon_{ijk}\partial^k \sigma^{(a)}, 
    ~~~ \mathbf{B}^{(a)}=\frac{g}{2\pi L}\nabla  \sigma^{(a)}.
\ee
In the last step of (\ref{topdensity}) we have replaced the field in the compact direction by it's vacuum expectation value since we are considering a semiclassical approximation.
The expression for the magnetic field  in terms of scalar magnetic potential should not be surprising as our system is in fact magnetostatic and a description in terms of $\sigma^{(a)}$ is quite appropriate to study the relevant dynamics. 

The explicit form for the creation operator for a monopole of type $a$ at $\mathbf{x}$ is given by \cite{Thomas:2011ee}
\be
\label{operator}
  {\cal{M}}_a (\mathbf{x}) =e^{i \alpha_{a} \cdot \bm{\sigma} (\mathbf{x})},
\ee
and for an antimonopole by
\be \label{antioperator}
  {\bar{\cal{M}}}_a (\mathbf{x})= e^{-i \alpha_{a} \cdot \bm{\sigma} (\mathbf{x})}.
\ee
The expectation values of these operators $\la {\cal{M}}_a (\mathbf{x}) \ra$ in fact determine the ground state of the theory.
Formula (\ref{operator}) shows again that $\bm{\sigma} (\mathbf{x})$ can be interpreted as a magnetic scalar potential.  
\exclude{
To be more precise, the effective magnetic potential $V_{mag} (\mathbf{x})$ is proportional to 
\be \label{potential}
  V_{mag} (\mathbf{x}) &\sim&   \bm{\sigma}  ~~~ \text{for} ~~    \bm{\sigma}   \in [0, \pi), \nonumber \\
  V_{mag} (\mathbf{x}) &\sim&  (2\pi- \bm{\sigma}) ~~~ \text{for}~~   \bm{\sigma}   \in [\pi, 2\pi)
\ee
as a result of periodic properties of the $\bm{\sigma} (\mathbf{x})$ field.
 as eqs. (\ref{thetaaction}), (\ref{operator}) state. }

Finally, the dimensional parameter which governs the dynamics of the problem is  defined as 
\be \label{sigmamass}
  m_{\sigma}^{2} \equiv L \zeta \left( \frac{2\pi}{g} \right)^{2}.
\ee
This parameter can be interpreted as Debye correlation length of the monopole's gas. The average number of monopoles in a ``Debye volume" is given by
\begin{equation} \label{debye}
  {\cal{N}}\equiv m_{\sigma}^{-3} \zeta = \left( \frac{g}{2\pi} \right)^{3} \frac{1}{\sqrt{L^3 \zeta}} \gg 1,
\end{equation} 
The above inequality holds since the monopole fugacity is exponentially suppressed, $\zeta \sim e^{-1/g^2}$, and we should view (\ref{debye}) as a constraint on the validity of the approximation where semiclassical approximation is justified. 

\section{Domain Walls in deformed QCD} \label{dw}

There is a discrete set of degenerate vacuum states as a result of the  $2\pi $ periodicity of the effective Lagrangian  (\ref{thetaaction}) for the $\bm{\sigma}$ field, and thus there exist domain wall configurations interpolating between these states.
The corresponding configurations are not however conventional domain walls similar to the well known ferromagnetic domain walls in condensed matter physics which interpolate between physically {\it distinct} vacuum states.
Here, instead, the corresponding configuration interpolates between topologically different but physically equivalent winding states $|n\ra$, which are connected to each other by large gauge transformation operator.
Therefore, the corresponding domain wall configurations in Euclidean space are interpreted as configurations describing tunnelling processes in Minkowski space, similar to Euclidean monopoles which also interpolate between topologically different, but physically identical states.
\exclude{
This interpretation should be contrasted with conventional interpretation of static domain walls defined in Minkowski space when corresponding solution interpolate between physically distinct states.
}
     
In fact, a similar domain wall which has an analogous interpretation is known to exist in QCD at large temperature in weak coupling regime where it can be described in terms of classical equation of motion.
These are so-called $Z_N$ domain walls which separate domains characterized by a different value for the Polyakov loop at high temperature.
As is known, see the review papers \cite{smilga,Fukushima:2011jc} and references therein, these $Z_N$ domain walls interpolate between topologically different but physically identical states connected by large gauge transformations similar to our case.
These objects can be described in terms of classical equation of motion and have finite tension $\sim T^3$ such that their contribution to path integral is strongly suppressed.
While the corresponding topological sectors are still present in the system at low temperature (though they are realized in a different way) it is not known how to describe the fate of $Z_N$ walls withcan become bound to the domain wallin QFT in the strong coupling regime wherein the semiclassical approximation breaks down. 
    
The domain walls to be discussed below in deformed QCD are very much the same as $Z_N$ domain walls at high temperature and their contribution to path integral is also strongly suppressed as their tension is finite in weak coupling regime. 
Nevertheless, one can study the structure of these domain walls, as well as their interaction   with dynamical magnetic monopoles.  
\exclude{Furthermore, as we discussed in Section \ref{introduction} the domain wall structure is apparently observed in the lattice simulations, which imply that they may have effectively vanishing tension at low temperature.}
We conjecture that the domain walls we describe below in the weak coupling regime in deformed QCD slowly become the objects  (with effectively vanishing tension) which are observed in lattice simulations \cite{Horvath:2003yj,Horvath:2005rv,Horvath:2005cv,Alexandru:2005bn,Ilgenfritz:2007xu,Ilgenfritz:2008ia,Bruckmann:2011ve,Kovalenko:2004xm} in the strong coupling regime, as we adiabatically increase the coupling constant without hitting the phase transition as argued in \cite{Yaffe:2008}.
This portion of the theory can not be tested in our deformed QCD model in the semiclassical approximation, but hopefully this portion of strongly coupled dynamics can be understood in the future using different techniques, such as the dual holographic description as advocated in the present context in \cite{Zhitnitsky:2011aa, BKYZ}.

\subsection{Domain wall solution}

There are many different types of domain walls supported by the system  (\ref{thetaaction})  which have very different physical meaning. 
In this paper we focus on the discrete symmetry of the effective Lagrangian (\ref{thetaaction}) given by the $2\pi$ shift, $\sigma_a \rightarrow  \sigma_a + 2\pi$, where any component of $\bm{\sigma}$ field can be shifted by $2\pi$ independently.
\exclude{
To simplify analysis, we consider a single specific non-vanishing component for the $\bm{\sigma}$ field sitting at $a-$th position, 
\be \label{ansatze}
  \bm{\sigma}  =  \left( 0, 0,  \sigma^{(a)}, 0, \dots, 0 \right)  ~~ a=1, ... N. 
\ee
This component describes a specific diagonal element of the original non-Abelian field strength. For example, $\chi^{(1)}$ corresponds to the following structure in conventional matrix notations
\begin{eqnarray} \label{matrix}
  \mathbf{B}^{(1)}=\frac{g}{2\pi L}\nabla  \sigma^{(1)}\cdot \left(
    \begin{array}{cccc}
      1 & 0 & ... & 0\\
      0 & -1 & ... & 0\\
      ... & ... & ... & 0\\
      0 & 0 & 0 & 0
    \end{array}
  \right) \, .
\end{eqnarray}
}
There are $N$ different domain wall types similar to monopole case since classification of our system is based on $\alpha_i\in \Delta_{\rm aff}$.
We emphasize that there are only $(N-1)$ physical propagating photons in the system as one scalar singlet field, though it remains massless, completely decouples from the system, and does not interact with other components at all \cite{Yaffe:2008}.
As a result of this structure, a configuration with $N$ different types of magnetic monopoles will carry zero magnetic charge and one unit of the topological charge $Q=1$ as each monopole carries $Q=1/N$ topological charge. 
The corresponding configuration can be  identified with a  conventional instanton with $Q=1$ which is made of $N$ constituents.
A similar comment also applies to DW structure: a configuration with $N$ different types of DWs on top of each other will produce a trivial vacuum configuration as the $(N-1)$ abelian components of the magnetic field will cancel each other, similar to magnetic monopole construction.
Thus, although there are $N$ different types of the DWs in our construction, only $(N-1)$ of them are independent.

\begin{figure}[t]
  \begin{center} 
    \includegraphics[width = 0.4\textwidth]{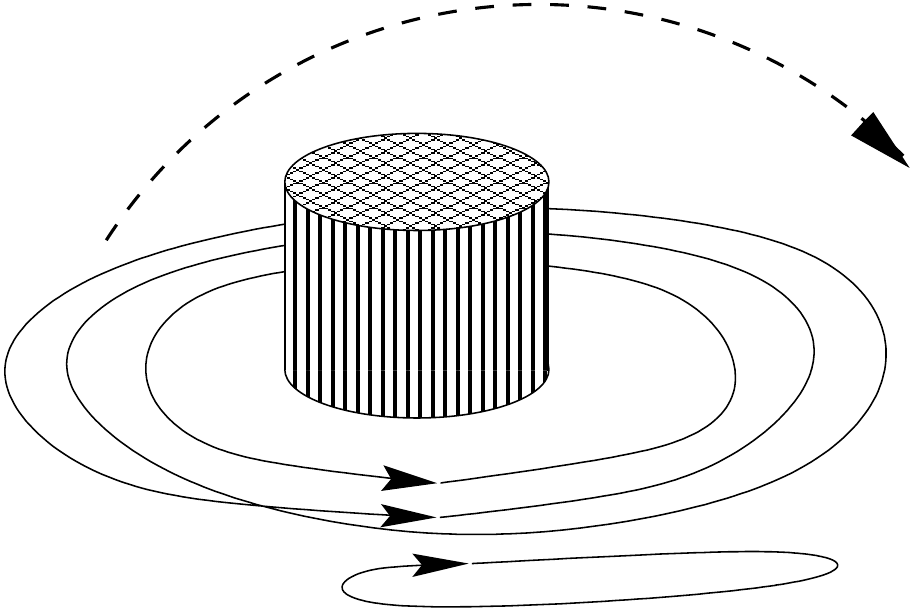}
      \caption{		\label{homotopy}
	This picture explains the transition between paths corresponding to the decay of some domain wall state to a domain wall free ground state.
	The path wrapping the peg represents a state with some domain walls, while the path that does not denotes a state with no domain walls
	We can deform the DW path by lifting it over the obstacle so that we can unwind it and deform it into the DW-free path.
	If the path describes domain walls with some weight, then it would require some energy to lift over the obstacle.
	If this energy is not available, then classically, the configurations that wind around the peg are stable.
	Quantum mechanically, however, the domain wall could still tunnel through the peg, and so the configurations are unstable quantum mechanically, see estimate for this probability in Appendix \ref{decay}. Picture adapted from \cite{Forbes:2000et}.
      }
  \end{center}
\end{figure}

In what follows, without loss of generality, we consider the $N=2$ case.
In this case there is only one physical field $\chi=(\sigma_1-\sigma_2)$ which corresponds to a single diagonal component from the original $SU(2)$ gauge group. The orthogonal combination $(\sigma_1 + \sigma_2)$ decouples from the system as explained in then original paper  \cite{Yaffe:2008}.
The  action (\ref{thetaaction}) becomes, 
\be	\label{action}
  S_{\chi}&=& \int_{\mathbb{R}^{3}}  d^{3}x \frac{1}{4 L} \left( \frac{g}{2 \pi} \right)^{2}
    \left( \nabla \chi \right)^{2} \\ &-&    \zeta  \int_{\mathbb{R}^{3}}  d^{3}x  \left[\cos \left( \chi 
    + \frac{\theta}{2} \right) +  \cos \left(- \chi + \frac{\theta}{2} \right) \right]. \nonumber
\ee
In terms of $\chi$ field, the classical equation of motion which follows from (\ref{action}) and which determines the profile of the domain wall has the form, 
\be \label{equation}
  \nabla^2 \chi-m_{\chi}^2\sin\chi=0,
\ee
where we take $\theta=0$ for simplicity, and the mass of $\chi$ field $m_{\chi}=2 m_{\sigma}$ is related to the Debye correlation length (\ref{sigmamass}).
The solution of this sine-Gordon equation which interpolates between $\chi(z=-\infty)=0$ and $\chi(z=+\infty)=2\pi$, and which is centered at $z_0=0$ being independent of $x,y$ coordinates is well known
\be \label{solution}
  \chi(z) = 4\arctan \left[\exp(m_{\chi}z)\right].
\ee
\exclude{We are now in position to explain the physical meaning of this solution.}
As we mentioned before, the domain wall (\ref{solution}) does not describe a physical domain wall (DW) which interpolates between physically {\it distinct} vacuum states, but rather interpolates between topologically different but physically {\it identical} states.
We remark that a similar construction has been considered previously in relation with the so-called $N=1$ axion model \cite{Vilenkin:1982ks,Hagmann:2000ja}, more recently in the QCD context in \cite{Forbes:2000et}, and in high density QCD in \cite{Son:2000fh}. 
In the previously considered cases \cite{Vilenkin:1982ks,Hagmann:2000ja,Forbes:2000et,Son:2000fh} as well as in present case (\ref{solution}) there is a single physical unique vacuum state, and interpolation (\ref{solution}) corresponds to the transition from one to the same physical state.
Therefore, such domain walls are not stable objects, but will decay quantum mechanically, see Appendix \ref{decay} for corresponding estimates.
Nevertheless, if life time of configuration (\ref{solution}) is sufficiently large, it can be treated as stable classical background, and it can be used to study the interaction of domain walls with monopoles, which is one of the main objectives of present work, see Figure \ref{homotopy} with more explanations. 

\exclude{
One can view these ``additional'' vacuum states which are physically {\it identical} states and which have extra $2\pi$ phase in operator (\ref{operator}) as an analog to the Aharonov Bohm effect with integer magnetic fluxes where electrons do not distinguish  integer fluxes from identically zero flux.
Our domain wall solution (\ref{solution}) describes interpolation between these two physically identical states.
}
Finally, one should also comment that, formally, a similar soliton-like solution which follows from the action (\ref{action}) appears in the computation of the string tension in Polyakov's 3d model \cite{Polyakov,Yaffe:2008}.
The solution considered there emerges as a result of the insertion of external sources in a course of computation of the vacuum expectation of the Wilson loop.
In contrast, in our case, the solution (\ref{solution}) is an internal part of the system without any external sources.
Furthermore, the physical meaning of these solutions are fundamentally different.
In our case the interpretation of the solution (\ref{solution}) is similar to an instanton describing the tunnelling processes in Minkowski space, while in the computations ~\cite{Polyakov,Yaffe:2008} it was an auxiliary object which appears in the course of computation of the string tension.

The width of the domain wall is determined by $m_{\chi}^{-1}$, while the domain wall tension $\sigma$ for profile (\ref{solution}) can be computed and it is given by
\be \label{tension}
  \sigma & = & 2\cdot \int^{+\infty}_{-\infty}   dz \frac{1}{4 L^2} \left( \frac{g}{2 \pi} \right)^{2}
    \left( \nabla\chi \right)^{2}\nonumber\\
   & = & \frac{m_{\chi}}{L^2} \left( \frac{g}{2 \pi} \right)^{2} \sim  \sqrt{\frac{\zeta}{L^3}}.
\ee

With explicit solution at hand (\ref{solution}), the magnetic field (\ref{magnetic}) distribution inside the domain wall is given by
\be \label{B}
  {B}_z=\left(\frac{g}{4\pi L}\right) \frac{4 m_{\chi}}{(e^{m_{\chi}z}+e^{-m_{\chi}z})},
\ee
The topological charge density distribution can then be computed using formula (\ref{topdensity}) with the following result
\be \label{Q}
  q(z)=\frac{\zeta}{L}\sin\chi=\frac{4\zeta}{L} 
    \frac{ (e^{m_{\chi}z}-e^{-m_{\chi}z}) }{(e^{m_{\chi}z}+e^{-m_{\chi}z})^2}.
\ee
 
\begin{figure}[t]
  \begin{center} 
    \includegraphics[width = 0.45\textwidth]{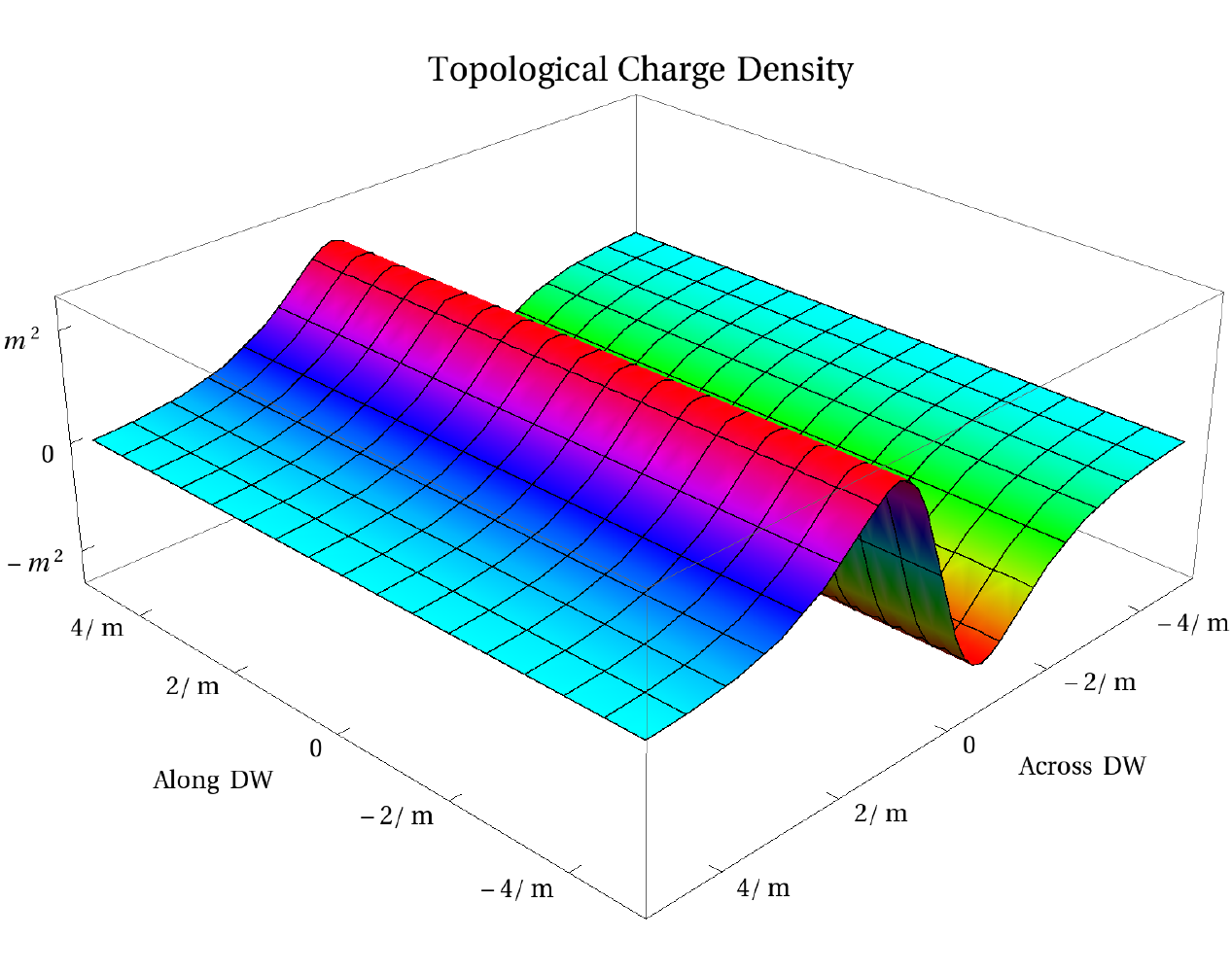}
      \caption{		\label{3DDomainWall}
	This picture shows the two layer structure of the topological charge density plotted against one direction across the Domain Wall and the other one of the two dimensions along it.
      }
  \end{center}
\end{figure}

From equation (\ref{Q}), we see that the net topological charge $Q\sim\int^{\infty}_{-\infty} dz q(z)$ on the domain wall vanishes.
However, the charge density has an interesting distribution; it is organized in a double layer structure, which is precisely what apparently has been measured  in the lattice simulations \cite{Horvath:2003yj,Horvath:2005rv,Horvath:2005cv,Alexandru:2005bn}.
For a graphical depiction see Figure \ref{3DDomainWall}.
The same double layer structure can be seen by computing the magnetic charge density $\rho_M$ which is defined as
\be \label{rho_M}
  \rho^{(a)}_M&\equiv& \left[ \nabla \cdot \mathbf{B}^{(a)} (\mathbf{x}) \right]= 
    \left(\frac{g}{4\pi L}\right)\frac{\partial^2\chi}{\partial z^2} \nonumber\\
  &=& 4\zeta \cdot \left(\frac{4\pi}{g}\right) \frac{ (e^{m_{\sigma}z}-e^{-m_{\sigma}z}) }
    {(e^{m_{\sigma}z}+e^{-m_{\sigma}z})^2}.
\ee
Thus, the relation between the topological charge density (\ref{Q}) and magnetic charge density (\ref{rho_M}) holds for the domain wall 
\be
  q(z)= \left(\frac{g}{2\pi }\right)\cdot \left(\frac{1}{LN}\right)\cdot \rho_M (z)
\ee
in full agreement with the general expression (\ref{topdensity}). 

From eqs. (\ref{Q}), (\ref{rho_M}), we see that an average density of magnetic monopoles filling the interior of domain wall is expressed in terms of  the same parameter  $\zeta$  which  characterizes  the average monopole's density in the system (\ref{thetaaction}).  One can interpret this relation as a hint that the topological charge sources have a tendency to reside in vicinity of the  domain walls rather than being uniformly distributed. We further elaborate on this matter in section \ref{dw-monopoles}. 
 
It is interesting to note that the domain walls in deformed QCD model are very similar (algebraically) to the well known domain wall studied previously in some SUSY models, see e.g. review \cite{Tong:2005un}.
Of course, there are fundamental differences between the two: in SUSY models the domain walls interpolate between physically distinct vacuum states, in huge contrast with our domain walls which correspond to interpolation between topologically different but physically identical states. 
Therefore, the interpretation in these two cases is fundamentally different: in SUSY models the domain walls are real physical objects, while in our deformed QCD  model they should be interpreted similar to instantons, objects which describe the tunnelling processes, see \cite{Zhitnitsky:2011aa} with more comments on this interpretation. 
Furthermore, the classification of the domain walls in SUSY models is based on flavour group symmetry breaking $SU(N_F)\rightarrow U(1)^{N_F-1}$, in contrast with colour symmetry breaking in deformed QCD.
However, the  formal classification of the domain walls in SUSY models based on simple roots from flavour group is very much  the same as classification in our case based on $SU(N)\rightarrow U(1)^{N-1}$ breaking pattern, see (\ref{roots}),(\ref{dotproduct}).
These similarities include, in particular, highly nontrivial properties such as ordering of the domain walls or their passing  through each other. However, these questions will not be elaborated on in the present work. 
  
The most important lesson from this analysis is that the double layer structure naturally emerges in the construction of the domain walls in the weak coupling regime in deformed QCD.
As claimed in \cite{Yaffe:2008} the transition from  high temperature weak coupling regime to low temperature strong coupling  regime should be smooth without any phase transitions on the way. 
Therefore, it would be tempting to identify the double layer structure found in this work (\ref{Q}) with the double layer structure from lattice measurements \cite{Horvath:2003yj,Horvath:2005rv,Horvath:2005cv,Alexandru:2005bn} when one slowly moves along a smooth path from the weak coupling to the strong coupling regime. 
 
\section{DW-Monopole Interaction}\label{dw-monopoles}

\exclude{
From expressions (\ref{operator}) and (\ref{ansatze}) one can infer that the interaction (to be more precise, the algebraic structure) of the DW with monopoles is very similar to monopole -monopole/anti monopole interactions. 
As our DWs are not dynamical configurations of the system, but rather, should be treated as a background classical fields we can not address the hard questions such as: what is the density of DW configurations, or why the topological charge density is mostly spread out in the domain walls rather than in localized objects as lattice simulations suggest \cite{Horvath:2003yj,Horvath:2005rv,Horvath:2005cv,Alexandru:2005bn}. 
Rather we can formulate a different question which can be addressed in weakly coupling regime.
What happens to the monopoles if they are formed in the presence of the domain walls?
As number of different types of domain walls, being $N$, equals to the number of different types of monopoles, being also $N$, one could assume that a specific monopole type $``a''$ will find a corresponding most attractive DW.
Therefore, we concentrate below on analysis of a specific configuration which contains two relevant elements: DW of type (\ref{matrix}) and a nearby anti-monopole with magnetic charge $-\alpha_1$ and topological charge $Q=-1/N$.
}

We now consider the domain wall configurations discussed in the previous section interacting with monopole configurations.
In these computations the domain walls are treated as classical background fields, and as such we do not consider fundamentally quantum questions, such as the density of domain walls.
Instead, we consider some questions which can be answered in the semiclassical context.
We focus on the interaction between a monopole and domain wall, each acting as magnetic sources, and compute the energy of the configuration as a function of separation distance between the two.
Therefore, the question we are addressing is where would a point charge prefer to sit in the presence of our domain wall? 
This question is actually motivated by lattice simulations  \cite{Horvath:2003yj,Horvath:2005rv,Horvath:2005cv,Alexandru:2005bn}
which suggest that the density of well localized 4d objects (such as small size 4d instantons) apparently vanishes, see item 6) from Introduction. 
We should emphasize that our domain walls are not empty objects as they are   already filled by magnetic monopoles with density determined by (\ref{Q}). 

Again, we consider the simplified scenario of $SU(2)$, which corresponds to considering the interaction between a single type of domain wall, $a$, and a monopole of the same type, or to be more precise an antimonopole so that the magnetic charge is $-\alpha_a$.
The domain wall is defined as previously, (\ref{solution}), but centered at a distance $z_0$ from the origin, so that the magnetic scalar potential is given by (letting $m = m_\chi$)
\be	\label{DWpot}
  \chi_{z_0}\left({\mb x}\right) = 4 \arctan \left[ e^{m \left( z - z_0 \right)} \right].
\ee
The monopole is defined such that it is a point source solution to the Klein-Gordon equation,
\be	\label{Klein-Gordon}
  \nabla_x^2 \varphi \left({\mb x }\right) - m^2 \varphi \left({\mb x }\right)
     = \delta \left({\mb x }\right),
\ee
centered at the origin ( ${\mb x_0} = 0$), and is thus an approximate solution to the sine-Gordon, (\ref{equation}), away from the origin.
The magnetic potential of the monopole is then given by the well known Yukawa potential,
\be	\label{Mpot}
  \varphi \left({\mb x}\right) = -\frac{e^{-m |{\mb x}|}}{4 \pi m |{\mb x}|}.
\ee

We then consider the configuration of monopole and domain wall separated by a distance $z_0$ and would like to compute the magnetostatic energy (Euclidean action) as a function of $z_0$. 
The energy associated with just a domain wall alone is proportional to the area of the domain wall, which is infinite in this case, so we compute instead the difference between the energy of the two together and the energy of the two independently,
\be	\label{computed}
  \Delta E \left( z_0 \right) = S \left[ \chi_{z_0} + \varphi \right]
     - S \left[ \chi_{z_0} \right] - S \left[ \varphi \right],
\ee
where $S$ is given by (axes have been rescaled relative to (\ref{action}))
\be	\label{simpleaction}
  S \left[ \chi \right] = \int_{\mathbb{R}^{3}}  d^{3}x \left[ \frac{1}{2} \left( \nabla \chi \right)^{2}
    - m^2 \cos \chi \right].
\ee
The quantity $\Delta E$ defines a ``binding energy'' and is finite. We cannot however compute it analytically, and so we compute above integrals numerically instead, for $z_0$ varying near the domain wall.
Some technical details of the computation are as follows.
We work in a cylindrical volume oriented across the domain wall such that it respects the symmetries of the physical geometry.
The cylinder is defined around the origin with radius $10/m$ and length $30/m$, so that we neglect the space outside of this region.
It is valid to do so since the monopole potential is exponentially suppressed with length constant $m$ and we are considering a bining energy.
We were forced to remove a small volume around the origin when computing the potential energy term because the structure is that of the cosine of a divergent quantity, which is highly oscillatory.
The potential energy due to the removed piece is bounded by the volume removed since it is a cosine so that we can make it arbitrarily small.
These two approximations make up the bulk of the numerical uncertainty, which is $\sim m^2/10^6$. 

\begin{figure}[t]
  \begin{center} 
    \includegraphics[width = 0.45\textwidth]{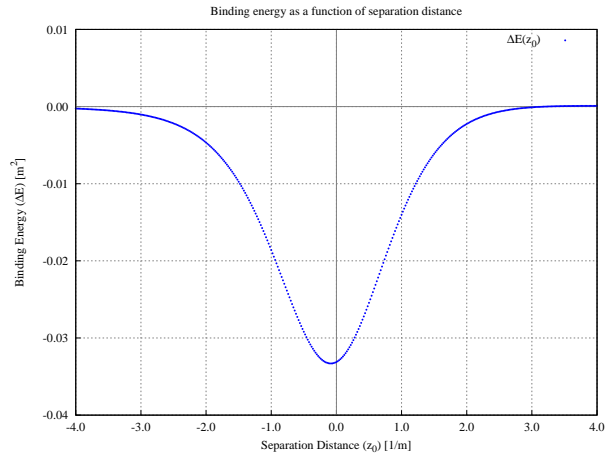}
      \caption{		\label{EnergyPositionPlot}
	This is a plot of the numerical result for the binding energy at various separation distances between domain wall and monopole.
	Notice that for $z_0 < 0$, the monopole to the right of the domain wall, there is an ``attractive'' potential with a minimum very near $z_0 = 0$.
      }
  \end{center}
\end{figure}

\begin{figure}[t]
  \begin{center} 
    \includegraphics[width = 0.45\textwidth]{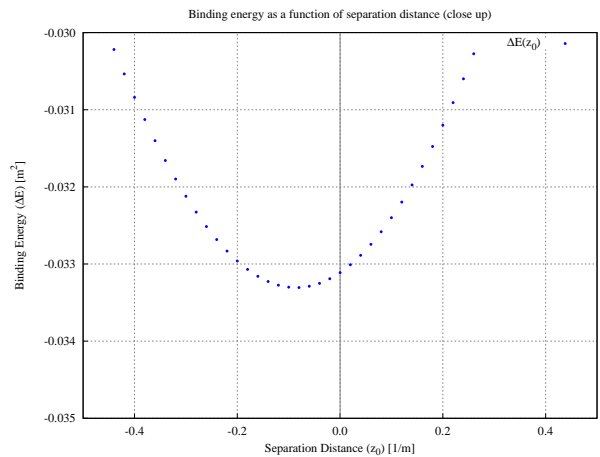}
      \caption{		\label{EPCloseUpCenterPlot}
	This plot is a close up of the points near the minimum in FIG. \ref{EnergyPositionPlot} showing that the minimum is slightly to the $z_0 < 0$ side of the center.
      }
  \end{center}
\end{figure}

\begin{figure}[t]
  \begin{center} 
    \includegraphics[width = 0.45\textwidth]{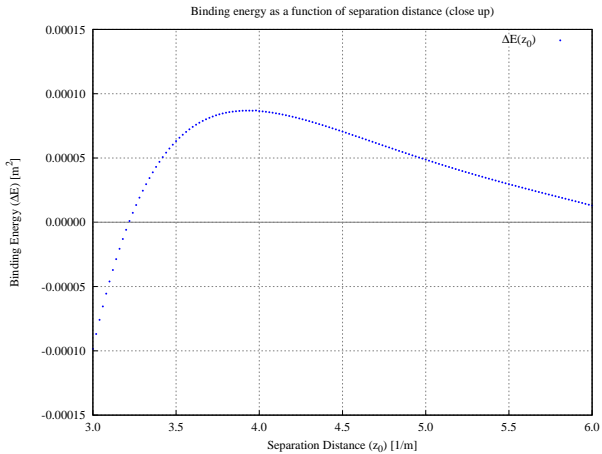}
      \caption{		\label{EPCloseUpRightPlot}
	This plot is a close up of the points to the right of FIG. \ref{EnergyPositionPlot} showing the small barrier present on the $z_0 > 0$ side. Notice the much finer vertical scale.
      }
  \end{center}
\end{figure}

Performing the numerical integration results in the plot given in Figure \ref{EnergyPositionPlot}.
There is an attractive potential between the monopole and domain wall with the monopole on one side ($z_0 < 0$), and a slightly repulsive one for the other side ($z_0 > 0$). The small barrier for $z_0 > 0$ is difficult to see in Figure \ref{EnergyPositionPlot} but obvious in Figure \ref{EPCloseUpRightPlot} which is just a plot of only points beyond $z_0 > 3$ with a much finer vertical scale.
Also, there is a minimum at $z_0 \sim 1/10m$ (see Figure \ref{EPCloseUpCenterPlot}), while the peak of the domain wall charge distribution is $\sim 1/m$. Thus the monopole would prefer to sit ``inside'' the domain wall, between the center and the peak of the sheet with the same charge density.
It is interesting that the monopole (with charge $-\alpha$) is attracted to the domain wall sheet with the same charge ($-\alpha$) rather than the sheet of opposing charge ($\alpha$), but the theory is non-linear so it is not altogether unexpected.

Figure \ref{EnergyPositionPlot} is not the complete story since we have not considered possible changes in the magnetic flux distribution coming from the monopole.
Basically, the monopole shape could deform in response to the interaction with the domain wall, so as to become less spherically symmetric.
In order to properly treat this problem, we should allow the spherical distribution of the monopole to vary to some superposition of solutions to the Klein-Gordon equation (\ref{Klein-Gordon}).
This described further calculation is beyond the scope of this work, but we conjecture that the magnetic field will prefer to orient itself along the domain wall, so that the magnetic flux will be pushed out to the edge of the domain wall at the boundary of space, similar to arguments presented in refs. \cite{Dvali:1996xe,Alexander:2000yx,Dvali:2002fi} in cosmological context. 
In this way, we have a picture in which any point-like magnetic monopoles become bound to extended domain walls with any magnetic flux being pushed along the domain walls to infinity. Apparently, this is precisely the picture discovered in lattice simulations \cite{Horvath:2003yj,Horvath:2005rv,Horvath:2005cv,Alexandru:2005bn} wherein very few localized 4d objects are observed in the system, see item 6) in the Introduction.

As a preliminary toward calculating the angular dependence if we allow the angular distribution to vary, we write a more general expression for a monopole-like solution to the Klein-Gordon equation (\ref{Klein-Gordon}), which depends on the angular coordinates:
\begin{equation} \label{GeneralKleinGordon}
  \varphi_n^m({\mb x}) \sim H^{(1)}_n(i m r)Y_n^m(\theta,\phi),
\end{equation}
where $H^{(1)}_n$ are the spherical Hankel functions of the first kind and the $Y_n^m$ are the spherical harmonics. Assuming the azimuthal axis is oriented across the domain wall, the problem is azimuthally symmetric and the spherical harmonics reduce to Legendre polynomials of $\cos(\theta)$.

When we attempt to calculate the binding energy as defined above for $\varphi_n$ it is negative and divergent for $n \ge 2$. It thus appears that the system is very sensitive to angular changes.
Furthermore, the divergence in $\Delta E$ seems to come from the core (near the divergence in $\varphi$) since it is highly sensitive to the amount of the core we remove when performing the numerical calculations. 
This however is also the region in which this approximation by Klein-Gordon monopoles is not really justified, and in fact the whole low-energy effective theory is suspect.
We therefore conclude that some other more sophisticated techniques will be required to address this problem of angular distribution, and as such it is well beyond the scope of this work.
Nevertheless, we do conjecture that the flux will have a tendency to spread along domain wall, but unfortunately cannot make a more quantitative claim at this point.

\section{Conclusion, Speculations, and Future directions}\label{conclusion} 

There are two important results of this work. Firstly, a double layer structure similar to that which is observed in lattice simulations \cite{Horvath:2003yj,Horvath:2005rv,Horvath:2005cv,Alexandru:2005bn} naturally emerges in the construction of the domain walls in weak coupling regime in deformed QCD.
Secondly, monopole configurations characterized by well localized topological (and magnetic) charge interact with domain walls in such a way that there is an attraction between the two, and the monopole favors a position inside the domain wall.
We introduced these domain walls as external background fields, while they are expected to  be dynamical configurations 
with effectively vanishing tension in strong coupling regime as holographic picture suggests. 
We further observe the tendency  that the magnetic field due a monopole in the presence of a domain will tend to align with the domain wall, such that the flux is pushed to the boundary of the domain wall.
If this effect persists in strongly coupled regime, it could be an explanation for the observation in lattice simulations \cite{Horvath:2003yj,Horvath:2005rv,Horvath:2005cv,Alexandru:2005bn} that there are no well localized objects with finite size which would carry the topological charge.

In weak coupling the domain wall solution is a nicely behaved smooth function, but what happens when we transition slowly to the strong coupling regime?
The holographic picture suggests that the effective domain wall tension vanishes and so they can be formed easily in vacuum.
It is possible that the domain walls become ``clumpy'' with a large number of folders.
Such fluctuations would then increase the entropy of the domain wall, which eventually could overcome the intrinsic tension. 
If this happens, the domain walls would look like very crumpled and wrinkled objects with large number of foldings, and as such, the domain walls may loose their natural dimensionality, and become characterized by a Hausdorff dimension as recent lattice simulations suggest \cite{Buividovich:2011cv}.
Nevertheless, the topological charge distribution on larger scales after averaging over a large number of these foldings should be sufficiently smooth so that the double layer structure would not disappear because the transition from weak to strong coupling regime should be sufficiently smooth as argued in \cite{Yaffe:2008}.
Therefore, we identify the double layer structure  found in this work (\ref{Q}) with the double layer structure from the lattice measurements \cite{Horvath:2003yj,Horvath:2005rv,Horvath:2005cv,Alexandru:2005bn}.
These particularities of the transition from weak to strong coupling are also interesting future questions, which will likely require an analysis beyond the semi-classical level. 

Apparently, the presence of such domain walls is a specific manifestation of the topological order characterizing this system as argued in \cite{Zhitnitsky:2013hs}. It could be different manifestations of this long range order such as the physical degeneracy of the   ground state  if the Euclidean space $\mathbb{R}^{3} $ in eq.(\ref{thetaaction}) is additionally compactified on a large torus, i.e.  $\mathbb{R}^{3}\rightarrow \mathbb{T}^{2} \times  \mathbb{R}^{1}$ as argued in \cite{Zhitnitsky:2013hs} in close analogy with topologically ordered condensed matter systems. 
It should be contrasted with behaviour of  a conventional gapped theory when  any variations of the boundary conditions at arbitrary large distances
can not so drastically  change the system. 

As a final remark, it is interesting to note that this kind of long range structure is apparently required to interpret an observed  local violation of ${\cal P}$ parity in heavy ion collisions in terms of large ${\cal P}$ odd domains with $\theta_{ind}\neq 0$ as argued in ~\cite{Zhitnitsky:2012ej}. 
The corresponding  ${\cal P}$ odd domains can be identified with interpolating long range $\eta'$ field which traces a pure glue configuration (\ref{Q}) studied in the present work.
Furthermore, this long range structure, if it persists at strong coupling regime, would justify a key assumption made in \cite{Kharzeev:2007tn} devoted to local ${\cal P}$ violation in heavy ion collisions. 
Only if $\theta_{ind}\neq 0$ is correlated on large scales is the effective Lagrangian approach of \cite{Kharzeev:2007tn} justified.
The deformed QCD model studied in this work explicitly shows how this long range structure could in principle emerge.

\section*{Acknowledgements}
ARZ is thankful to  Gokce~Basar, Dima Kharzeev, Ho-Ung Yee,  Edward Shuryak  and other  participants of the workshop ``P-and CP-odd effects in hot and dense matter", Brookhaven, June, 2012, for useful and stimulating discussions related to the subject of the present work.
This research was supported in part by the Natural Sciences and Engineering Research Council of Canada.

\appendix
         
\section{Decay of domain walls}\label{decay}

The decay mechanism is due to a tunnelling process which creates a hole in the domain wall which connects the $\chi=0$ domain on one side of the wall to the $\chi = 2\pi$  domain on the other, see (\ref{solution}).
Because the ground state on the two sides is physically identical, it is possible for the fields to remain in the ground state as they pass through the hole.
That is, there is no interpolation (winding) as they pass through the hole.
This lowers the energy of the configuration over that where the hole was filled by the domain wall transition by an amount proportional to $R^2$  where $R$ is the radius of the hole.
The hole, however, must be surrounded by a string-like field configuration which interpolates between an unwound configuration and a wound one.
This string represents an excitation in the heavy degrees of freedom and thus costs energy, however, this energy scales linearly as $R$.
Thus, if a large enough hole can form, it will be stable and the hole will expand and consume the wall.
This process is commonly called quantum nucleation and is similar to the decay of a metastable wall bounded by strings; therefore, we use a similar technique to estimate the tunnelling probability.
The idea of the calculation was suggested in \cite{Vilenkin:1982ks} to estimate the decay rate in the so-called $N=1$ axion model.
In QCD context similar estimations have been discussed for the $\eta'$ domain wall in large $N$ QCD in \cite{Forbes:2000et} and for the $\eta'$ domain wall in high density QCD in \cite{Son:2000fh}. 
         
If the radius of the nucleating hole is much greater than the wall thickness, we can use the thin-string and thin-wall approximation.
This approximation justified as we shall see when we calculate the critical radius $R_c$.
In this case, the action for the string and for the wall are proportional to the corresponding worldsheet areas 
\be	\label{S_0} 
  S_0  \left(\mathbb{R}^{3}\times \mathbb{S}^{1}\right)=2\pi R L \alpha - \pi  R^2L \sigma .
\ee
The first term is the energy cost of forming a string, where $\alpha$ is the string tension and $2\pi R L$  is its worldsheet area. 
The second term is energy gain by the hole over the domain wall, in which $\sigma$ is the wall tension and $\pi  R^2L$ is its worldsheet volume. 
We should note  that formula (\ref{S_0}) replaces following, more familiar expression for the classical action which was used  in many previous similar computations, see \cite{Forbes:2000et,Son:2000fh}
\be	\label{S_1}
  S_0 (\mathbb{R}^{4})=4\pi R^2 \alpha -\frac{4\pi}{3} R^3 \sigma .
\ee
Minimizing (\ref{S_0}) with respect to $R$ we find the critical radius $R_c$ and the action $S_0$ 
\be	\label{R}
  R_c=\frac{\alpha}{\sigma}, ~~S_0 \left(\mathbb{R}^{3}\times \mathbb{S}^{1}\right)= \frac{\pi\alpha^2 L}{\sigma}, 
\ee
which replace  more familiar expressions for the critical radius $R_c=\frac{2\alpha}{\sigma} $ and classical action $S_0 (\mathbb{R}^{4})= \frac{16\pi\alpha^3}{3\sigma^2}$ from \cite{Forbes:2000et,Son:2000fh}.
        
Therefore, the semiclassical probability of this process is proportional to
\be	\label{decay_rate}
  \Gamma \sim \exp \left(-\frac{\pi\alpha^2 L}{\sigma}\right)
\ee
where $\sigma $ is the DW tension determined by (\ref{tension}), while $\alpha$ is the tension of the vortex line in the limit  when the interaction term $\sim \zeta$ due to the monopole's interaction in low energy description (\ref{thetaaction}) is neglected and $U(1)$ symmetry is restored.
In this case the vortex line is a global string with logarithmically divergent tension
\be	\label{vortex}
  \alpha\sim 2\pi\frac{1}{4L^2} \left( \frac{g}{2 \pi} \right)^{2}\ln \frac{R}{R_{core}}
\ee
where $R\sim m_{\chi}^{-1}$ is a long-distance cutoff which is determined by the width of the domain wall, while $R_{core}\sim L$ when low energy description breaks down.
The vortex tension is dominated by the region outside the core, so our estimates for computing $\alpha$ to the logarithmic accuracy are justified.
Furthermore, the critical radius can be estimated as 
\be	\label{R_c}
  R_c=\frac{\alpha}{\sigma}\sim \frac{\pi}{2m_{\chi}}\ln(\frac{1}{m_{\chi}L} ),
\ee
which shows that the nucleating hole $\sim R_c$ is marginally greater than the wall thickness $\sim m_{\chi}^{-1}$ as logarithmic factor $\ln(\frac{1}{m_{\chi}L}) \sim \ln {\cal{N}}\gg 1$  where ${\cal{N}}\gg 1$ is large parameter of the model, see (\ref{debye}).
Therefore, our thin-string and thin-wall approximation is marginally justified. 
                  
As a result of our estimates (\ref{decay_rate}), (\ref{tension}), (\ref{vortex}) the final expression for the decay rate of the domain wall is proportional to
\be	\label{rate}
  \Gamma &\sim& \exp \left(-\frac{\pi\alpha^2 L}{\sigma}\right)
    \sim \exp \left(- {\pi^3}\left(\frac{g}{4\pi}\right)^3
    \frac{ \ln^2  (\frac{1}{m_{\chi}L} )}{\sqrt{L^3\zeta}}\right) \nonumber\\
  &\sim& \exp \left(-\gamma\cdot {\cal{N}}\ln^2 {\cal{N}} \right) \ll 1,
\ee
with $\gamma$ being some numerical coefficient.
The estimate (\ref{rate}) supports our claim that in deformed QCD model when weak coupling regime is enforced  and ${\cal{N}}\gg 1$ the domain walls are stable objects and our treatment of the domain walls as stable objects is justified.

\end{document}